\documentclass[11pt,letterpaper]{article}

\usepackage{fullpage}
\usepackage[numbers]{natbib}

\usepackage{booktabs}
\usepackage{wrapfig}
\usepackage{graphicx}
\usepackage{caption,subcaption}

\usepackage[ruled]{algorithm2e} 

\newcommand{\ignore}[1]{}

\usepackage{hyperref}

\newcommand{\PEER}{Peer Communication\xspace}

\newcommand{\peer}{peer communication\xspace}

\newcommand{\squishlist}{
   \begin{list}{{{\small{$\bullet$}}}}
    { \setlength{\itemsep}{0pt}      \setlength{\parsep}{1pt}
      \setlength{\topsep}{1pt}       \setlength{\partopsep}{0pt}
     \setlength{\leftmargin}{1em} \setlength{\labelwidth}{1em}
      \setlength{\labelsep}{0.5em} } }
\newcommand{\squishend}{  \end{list}  }

\begin{document}

\title{Working in Pairs: Understanding the Effects of Worker Interactions in Crowdwork}
\author{Chien-Ju Ho%
\footnote{Washington University in St. Louis. Email: {\tt chienju.ho@wustl.edu}.}
\and
Ming Yin
\footnote{Purdue University. Email: {\tt mingyin@purdue.edu}.}
\thanks{The authors are ordered alphabetically.}
}
\date{}

\maketitle

\begin{abstract}
Crowdsourcing has gained popularity as a tool to harness human brain power to help solve problems that are difficult for computers.
Previous work in crowdsourcing often assumes that workers complete crowdwork \emph{independently}.
In this paper, we relax the independent property of crowdwork and explore how introducing direct, synchronous, and free-style interactions between workers would affect crowdwork.
In particular, motivated by the concept of peer instruction in educational settings, we study the effects of {\em \peer} in crowdsourcing environments.
In the crowdsourcing setting with \peer, pairs of workers are asked to complete the same task together by first generating their initial answers
to the task independently and then freely discussing the tasks with each other and updating their answers after the discussion.
We experimentally examine the effects of \peer in crowdwork on various common types of tasks on crowdsourcing platforms, including image labeling, optical character recognition (OCR), audio transcription, and nutrition analysis.
Our experiment results show that the work quality is significantly improved in tasks with \peer compared to tasks where workers complete the work independently.
However, participating in tasks with \peer has limited effects on influencing worker's independent performance in tasks of the same type in the future.

\end{abstract}

\section{Introduction}
Crowdsourcing is a paradigm for utilizing human intelligence to help solve problems that computers alone can not yet solve. 
In recent years, crowdsourcing has gained increasing popularity as the Internet makes it easy to engage the crowd to work together.
On a typical crowdsourcing platform like Amazon Mechanical Turk (MTurk), 
task requesters may post ``microtasks'' that workers can complete \emph{independently} in a few minutes in exchange for a small amount of payment. 
A microtask might involve labeling an image, transcribing an audio clip, 
or determining whether a website is offensive.
Much of the practice and research in crowdsourcing has made this independence assumption and has focused on designing effective aggregation methods~\cite{raykar10,cholleti08,whitehill09} or incentive mechanisms~\cite{Mason:09,Ho15} to improve the quality of crowdwork.

More recently, researchers have started to explore the possibility of removing this independence assumption and enable worker collaboration in crowdsourcing.
One typical approach is to design {\em workflows} that coordinate crowd workers for solving complex tasks. 
Specifically, a workflow involves decomposing a complex task into multiple simple microtasks, and workers are then asked to work on {\em different} microtasks. 
Since decomposed microtasks may depend on each other (e.g., the output of one task may be used as the input for another),
workers are implicitly interacting with one another and are not working independently.
Along this line, there has been a great amount of research demonstrating that relaxing the worker 
independence assumption could enable us to go beyond microtasks and solve various complex tasks using crowdsourcing~\cite{Bigham10,Kittur:11,Kulkarni:12,Retelny:14}.

Another line of research has demonstrated that even when workers are working on the {\em same} microtask, 
enabling some form of {\em structured interactions} between workers could be beneficial as well. 
In particular, ~\citet{Drapeau16} and ~\citet{Chang17revolt} has shown that, in labeling tasks,
if workers are presented with alternative answers and the associated arguments, 
which are generated by other workers working on the same tasks,
they can provide labels with higher accuracy. 
These results, again, imply that including worker interactions could have positive impacts on crowdwork.

In both these lines of research, however, interactions between workers are {\em indirect} and {\em constrained} by the particular format of 
information exchange that is pre-defined by requesters (e.g., the input-output handoffs in workflows, the elicitation of arguments for workers' answers).
Such form of worker interactions can be context-specific and may not be easily adapted to different contexts. For example, it is unclear whether
presenting alternative answers and arguments would still improve worker performance for tasks other than labeling, where it can be hard for workers to provide a simple
justification for their answers. 


Naturally, one may ask what if we can introduce {\em direct}, {\em synchronous} and {\em free-style} worker interactions in crowdwork?
We refer to this alternative type of worker interactions as {\em \peer}, 
and in this paper, we focus on understanding the effects of \peer in crowdwork when a pair of workers are working on {\em the same microtask}.
In particular, inspired by the concept of peer instruction in educational settings~\cite{Crouch10},
we operationalize \peer as a procedure where a pair of workers working on the same task are asked to first provide an independent answer each, then freely discuss the
task, and finally provide an updated answer after the discussion. 
We ask the following two questions to understand the effects of \peer:

\begin{itemize}
    \item \textbf{Whether and why \peer improves the quality of crowdwork?}\\
    Empirical study on the effects of peer instruction suggests that students are more likely to provide correct answers to test questions after discussing with their peers~\cite{Crouch10}. 
    Moreover, previous work in crowdsourcing also demonstrates that indirect worker interactions (e.g., showing workers the arguments from other workers)~\cite{Drapeau16,Chang17revolt} improve the quality of crowdwork for labeling tasks. 
    We are thus interested in exploring whether \peer, a more general form of worker interactions, could also have positive impacts on the quality of crowdwork for a more diverse set of tasks.\\

    \item \textbf{Can \peer be used to train workers so that workers can achieve better independent performance on the same type of tasks in the future?}\\
	It is observed that students learning with peer instruction obtain higher grades when they (independently) take the post-tests at the the end of the semester~\cite{Crouch10}.
	Moreover, previous work in crowdsourcing also shows that some types of indirect worker interactions (e.g., asking workers to review or verify the results of other workers in the same type of task) could enhance workers' independent performance for similar tasks in the future~\cite{Doroudi:16,Zhu:14}.
	We are thus interested in examining whether \peer could also be an effective approach to train workers.
\end{itemize}

We design and conduct experiments on Amazon Mechanical Turk to answer these questions. 
In our first set of experiments,
we examine the effects of \peer with three of the mostly commonly seen tasks in crowdsourcing markets: image labeling, optimal character recognition, and audio transcriptions.
Experiment results show that workers in tasks with \peer perform significantly better than workers who work independently. 
The results are robust and consistently observed for all three types of tasks.
By looking into the logs of worker discussion, we find that  
most workers are engaged in constructive conversations and exchanging information that their peer might not notice or do not know. 
This observation reinforces our beliefs that consistent quality improvements can be obtained through introducing \peer in crowdwork.
However, unlike in the educational setting, workers who have completed tasks with \peer do not produce independent work of higher quality on the same type of tasks in the future.

We then conduct a second set of experiments with nutrition analysis tasks to examine the effects of \peer in training workers for future tasks in more depth.
The experiment results 
suggest that workers' independent performance in future tasks \emph{only} improves 
when the future tasks share {\em related} concepts to the training tasks (i.e., the tasks where \peer happens), and 
when workers are given {\em expert feedback} after \peer. 
Moreover, such improvement is likely caused by expert feedback rather than the \peer procedure.
In other words, we find that \peer, per se, may have limited effectiveness in training workers towards better independent performance, at least for microtasks in crowdsourcing.

Our current study focuses on one-to-one communication between workers on microtasks. 
We believe our results provide implications for the potential benefits of introducing direct interactions among multiple workers in complex and more general tasks,
and we hope more experimental research will be conducted in the future to carefully understand the effects of \peer in various crowdsourcing contexts.

\subsection{Related Work}
A major line of research in crowdsourcing is to design effective quality control methods.
Most of the work in this line has made the assumption that workers independently complete the tasks.
One theme in the quality control literature is the development of statistical inference and probabilistic modeling methods for the purpose of aggregating workers' answers. 
Assuming a batch of noisy inputs, the EM algorithm~\cite{dempster77} can be adopted to learn the skill level of workers and obtain estimates of the best answer~\cite{raykar10,cholleti08,jin03,whitehill09,dawid79}. 
There have also been extensions to also consider task assignments in the context of these probabilistic models of workers~\cite{KOS11,karger11a,HJV13}. 
Another theme is to design incentive mechanisms to motivate workers to contribute high-quality work.
Incentives that researchers have studied include monetary payments~\cite{Mason:09,Horton:10b,Yin:13,Ho15} and
intrinsic motivation~\cite{Law16,Rogstadius:11,Shaw:2011}.
In addition, gamification~\cite{Ahn06}, badges~\cite{Anderson13}, and virtual points~\cite{jain2009designing} are also explored to steer workers' behavior.

The goal of this work is to explore the effects of worker interactions in crowdsourcing environments. 
Researchers has explored implicit worker interactions through the design of workflows to coordinate multiple workers.
In a workflow, a task is decomposed into multiple microtasks, which often depend on each other, e.g., the output of one microtask is served as the input for another microtask.
As a result, workers are implicitly interacting with each other. 
For example, \citet{Little10} propose the Improve-and-Vote workflow, in which some workers are working on improving the current answer while other workers can vote on whether the updated answers are better than the original ones. \citet{Dai13} apply partially-observable Markov Decision Process (POMDP) to better set the parameters in these workflows (e.g., how many workers should be involved in voting). 
More complicated workflows have also been proposed to solve complex tasks using a team of crowdworkers~\cite{Noronha:11,Kittur:11,Kulkarni:12,Retelny:14}.
These workflow-based approaches enable crowdsourcing to solve not only microtasks but also more complex tasks.
However, worker interactions in this approach are often implicit and constrained (e.g., through the input-output handoffs).
We are interested in studying the effects of direct and free-style communications between workers in crowdwork.

Our work focuses on worker interactions when workers work on the same microtask and is related to that of~\citet{Drapeau16} and \citet{Chang17revolt}. 
\citet{Drapeau16} propose an Assess-Justify-Reconsider workflow for labeling tasks: given a labeling task, workers first assess the task and give their answers independently; workers are then asked to come up with arguments to justify their answers; finally, workers are presented with arguments from a different answer and are then asked to reconsider their answers.
They show that applying this workflow greatly improves the quality of answers generated by crowd workers.
\citet{Chang17revolt} also propose a similar Vote-Explain-Categorize workflow with an additional goal of collecting useful arguments as the labeling guidelines for future workers. 
Both these studies have relaxed the independence assumption and enabled worker interactions through presenting the arguments from another worker. 
However, they focus only on classification tasks (e.g., answering whether there is a cat in the image), and the worker interactions are limited to presenting arguments from another worker.
In this work, we are interested in enabling more general form of interactions (i.e., direct, synchronous, and free-style communications) for more diverse types of tasks.
In particular, in addition to classification tasks, we have explored worker interactions on optical character recognition and audio transcription. 
It is not trivial how the above two workflows can be applied in these tasks, as workers might not know how to generate arguments for these tasks without interacting with fellow workers in real time. 

Regarding the role of worker interactions in ``training'' workers, previous research~\cite{Zhu:14,Doroudi:16} suggests that, for complex tasks, introducing limited form of implicit worker interactions, e.g., providing (expert or peer) feedback to workers after they complete the tasks or 
asking workers to review or verify the work produced by other workers,
could improve workers' performance in the future. 
In this work, we focus on examining whether direct, synchronous, and free-style communication (instead of one-directional feedback or reviewing) can
be an effective training method to improve workers' independent performance in microtasks.
\ignore{
We examine the effects of \peer on both 1) tasks workers are working on during communications and 2) tasks workers will independently work on in the future.
~\citet{Koriat360} have shown that, for perceptual judgements tasks (e.g., deciding which shaded area is larger for a given pair of shapes), 
when users are can freely communicate with each other and exchange their confidence,
they would outperform users who work independently.
In this work, we are interested in understanding whether the results generalize to real tasks in crowdwork when the communication is guided by peer instruction.
\citet{Zhu:14,Doroudi:16} has shown that, for complex tasks, introducing limited form of worker interactions, e.g., providing (expert or peer) feedback to workers after they complete the tasks or giving workers access to others' work, could improve workers' performance in the future. 
In this work, we focus on microtasks and the effects of direct and freestyle communications (instead of one-directional feedback).
}

\citet{Niculae16} has explored whether interactions can help improve workers' output. 
They designed an online game in which online players can discuss together to identify the location where a given photo is taken. 
However, their focus is on applying natural language processing techniques to predict whether a discussion would be constructive based on analyzing users' chat logs. 
Their results could be useful and interesting to apply in our setting.

This work adopts the techniques from peer instruction, which is a widely adopted interactive learning approach in many institutions and disciplines~\cite{Crouch10,Fagen02peerinstruction,Lasry08peerinstruction,Porter11peer,Mazur17peer},
and has been empirically shown to better engage students and also help students achieve better learning performance. 
We will provide more details on the concept of peer instruction in the next section.

\section{\PEER in Crowdwork}
In this section, we give a brief introduction to the concept of peer instruction. 
We then describe our approach of \peer, which adapts peer instruction to crowdsourcing environments.

\subsection{Peer Instruction in Educational Settings}
Peer instruction is an interactive learning method developed by Eric Mazur which aims at engaging students for more effective learning during classes. 
Different from traditional teaching methods, which is typically centered around the instructors as the instructors convey knowledge to students in a one-sided manner through pure lectures, peer instruction creates a student-centered learning environment where students can instruct and learn from each other.

\begin{wrapfigure}{R}{0.45\textwidth}
   \centering
   \includegraphics[width=0.4\columnwidth, keepaspectratio]{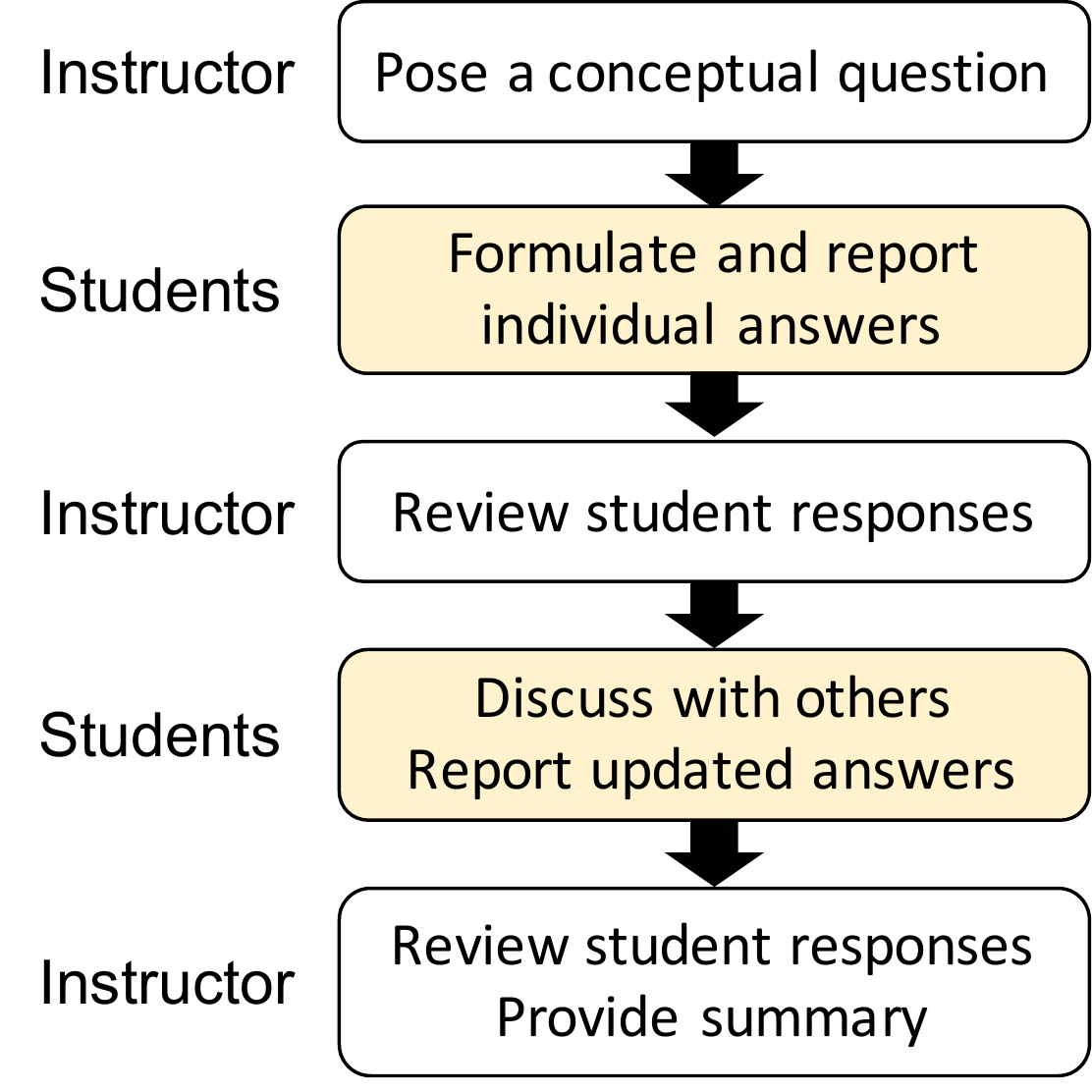}
   \caption{Questioning procedure of peer instruction.}
   \vspace{-10pt}
   \label{fig:PI-workflow}
\end{wrapfigure}

More specifically, peer instruction involves students first learning outside of class by completing pre-class readings and then learning in class by engaging in the conceptual question answering process. 
Figure~\ref{fig:PI-workflow} summarizes the in-class questioning procedure of peer instruction. 
Such procedure starts with the instructor proposing to students a question that is related to one concept in the pre-class readings.
Students are then asked to reflect on the question, formulate answers on their own, and report their answers to the instructor.
Next, students can discuss the question with their fellow students\footnote{In practice, if most of the students answer the question correctly on their own, the instructor can decide to skip the discussion phase and move on to the next concept.}.
During the discussion, students are encouraged to articulate the underlying reasoning of their answers and convince their peers that their answers are correct.
After the discussion, each student reports to the instructor a (final) updated answer, which may or may not be different from her initial answer before the discussion. 
Lastly, after reviewing students' final responses, the instructor decides to either provide more explanation on the concept associated with the current question or move on to the next concept.

\ignore{Intuitively, peer instruction turns students from passive listeners and knowledge consumers in class to active learners and de facto knowledge producers. 
Importantly, compared to instructors, students might be able to provide explanations that are better understood by other students as they share similar backgrounds.
For instructors, peer instruction also helps them to monitor students' learning progress by requesting students to provide feedback (through answering conceptual questions) during lectures. 
This extra information is valuable for instructors to adjust the pace of teaching and/or take additional interventions as needed.
Given these potential benefits, peer instruction has been used in a large number of institutions and disciplines~\cite{Crouch10,Fagen02peerinstruction,Lasry08peerinstruction}. }

The peer instruction method has been widely adopted in a large number of institutions and disciplines~\cite{Crouch10,Fagen02peerinstruction,Lasry08peerinstruction}.
Intuitively, peer instruction may improve learning as students become active knowledge producers instead of passive knowledge consumers. 
Compared to instructors, students might be able to provide explanations that are better understood by other students as they share similar backgrounds.
Empirical observations for deploying peer instruction confirm that it successfully improves students' learning performance~\cite{Crouch10}.
In particular, students are more likely to provide correct answers to the conceptual question after discussing peers than they do before the discussion. 
Moreover, in post-tests where students independently answer a set of test questions after the end of the semester,
students who participate in courses taught with peer instruction perform significantly better than students who don't.
These empirical evidences suggest that peer instruction helps students understand not only the current question but also the underlying concepts, which eventually help them obtain better independent performance in future tests.

\subsection{\PEER: Adapting Peer Instruction to Crowdwork}
\label{sec:pi}

We propose to study the effects of \peer in crowdwork, applying the idea of peer instruction as a principled approach to structure the direct interactions among crowd workers. 
In particular, given a particular microtask, we consider the requester of it as the ``instructor,'' all workers working on it as ``students,'' 
and the task per se as the ``conceptual question'' proposed by the instructor.   
Hence, a natural way to adapt peer instruction to crowdsourcing would be asking each worker working on the same task to first complete
the task independently and then allowing them to discuss the task with each other and submitting their final answers. 

The success of peer instruction in improving students' learning performance in the educational domain implies the possibility
of using such strategy to enhance the quality of work in crowdsourcing, both on tasks where \peer takes place and on future tasks of the same type.
However, it is unclear whether the empirical findings
on the effects of peer instruction in educational settings can be directly generalized to the crowdsourcing domain. 
For example, while conceptual questions in educational settings typically involve problems that require specialized knowledge or domain expertise,
crowdwork is often composed of ``microtasks'' that only ask for simple skills or basic intelligence. 
Moreover, in peer instruction, the instructor can provide additional explanations to clarify confusions students might have during the discussion.
However, in \peer, requesters often do not know the ground truth of the tasks and are not able to provide additional feedback after workers submit their tasks.

Therefore, in this work, we aim to examine the effects of \peer in crowdsourcing, and in particular, whether \peer has positive effects on the quality of crowdwork.
More specifically, based on the empirical evidence on the effectiveness of peer instruction in education as well as the positive effects of indirect worker interactions demonstrated
in previous research, we postulate two hypotheses on the effects of \peer in crowdwork:

\begin{itemize}
    \item \textbf{Hypothesis 1 (H1)}: Workers can produce higher work quality in tasks with \peer than that in tasks where they work independently.
    \item \textbf{Hypothesis 2 (H2)}: 
    After \peer, workers are able to produce independent work of higher quality on the same type of tasks in the future.
\end{itemize}

We design and conduct a series of large-scale online experiments to test these two hypotheses. 
In our experiments, we operationalize the procedure of 
\peer between {\em pairs} of workers who are working on the same microtask, and we leave the examination of the effects of \peer in larger groups of crowd workers on more complex tasks as future work. 
It is also worthwhile to note that, in this work, we focus on adapting the component of worker interactions in peer instruction (i.e., the yellow-shaded boxes in Figure~\ref{fig:PI-workflow}).
However, in practice, the requester can often make interventions to improve the efficiency of the \peer process (e.g., given the initial answers submitted before discussion, the requester can dynamically decide how to present information or match workers to make the discussions more effective). 
The study of requester interventions in \peer is out of the scope of the current paper, but it is another direction that worths further research.

\section{Experiment 1: How Does Peer Communication Affect Quality of Crowdwork?}
To examine how introducing direct communication between pairs of workers in crowdwork affects the work quality, we design and conduct a set of online experiments on Amazon Mechanical Turk (MTurk) with three types of microtasks that commonly appear on crowdsourcing platforms, including image labeling, optical character recognition (OCR), and audio transcription.


\subsection{Independent Tasks vs. Discussion Tasks}

As previously stated in our hypotheses, we are interested in understanding whether allowing workers to work in pairs and directly communicate with each other about the {\em same} tasks would lead to work of better quality 
compared to that when workers complete the work independently, both on tasks where peer communication happens (H1) and on future tasks of the same type after \peer takes places (H2). To do so, in our experiments, we consider both tasks {\em with} peer communication and tasks {\em without} peer communication:

\begin{itemize}
    \item \textbf{Independent tasks} (tasks without peer communication). 
    In an independent task, workers are instructed to complete the task on their own. 

    \item \textbf{Discussion tasks} (tasks with \peer). 
    Workers in a discussion task are guided to communicate with other workers to complete the task together, following a process adapted from the peer instruction
    procedure as we have discussed in Section~\ref{sec:pi}. In particular, each worker is paired with another ``co-worker'' on a discussion task. 
    Both workers in the pair are first asked to work on the task independently and submit their independent answers.
    Then, the pair of workers enter a chat room, where they can see each other's independent answer to the task, and they are given two minutes to discuss the task freely. 
    Workers are instructed to explain to each other why they believe their answers are correct. 
    After the discussion, both workers get the opportunity to update
    their answers and submit their final answers.
\end{itemize}

\begin{figure*}[t]
   \centering
   \includegraphics[width=1\columnwidth, keepaspectratio]{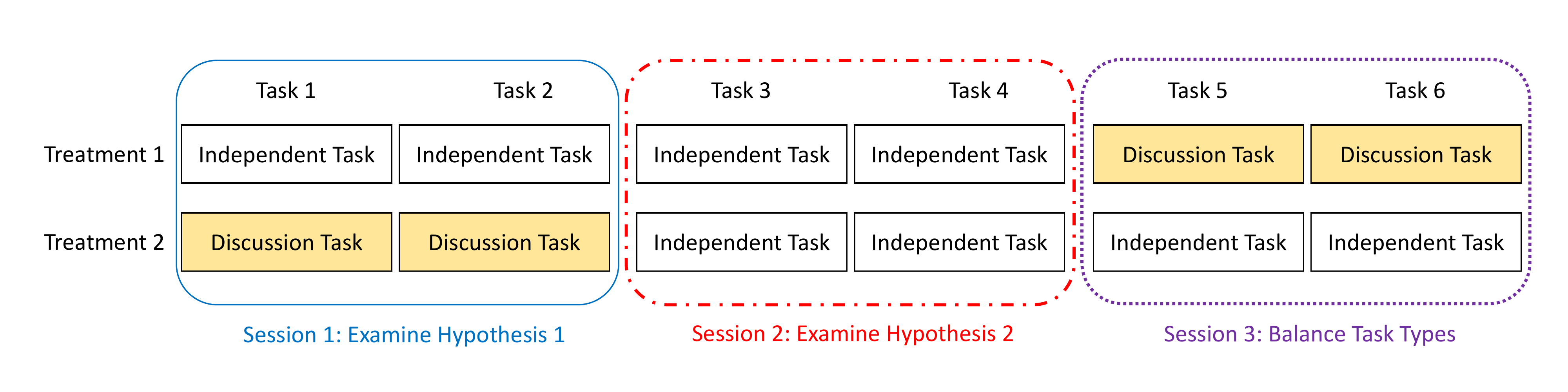}
   \vspace{-20pt}
   \caption{The two treatments used in our experiments. 
        This design enables us to examine Hypothesis 1 (through comparing work quality in Session 1) and Hypothesis 2 (through comparing work quality in Session 2), while not creating significant differences between the two treatments (through adding Session 3 to make the two treatments containing equal number of independent tasks and discussion tasks).
        }
   \label{fig:exp-design}
   \vspace{-10pt}
\end{figure*}

\subsection{Treatments}
\label{two-treatment-design}

We conduct randomized experiments to examine our hypotheses regarding the effects of \peer on the quality of crowdwork.
The most straight-forward experimental design would include two treatments, where workers in one treatment are asked to work on a sequence of 
independent tasks while workers in the other treatment complete a sequence of discussion tasks.
However, since the structure of independent tasks and discussion tasks are fundamentally different---discussion tasks naturally require more time and effort from workers but
can be more interesting to workers---it is possible for us to observe significant self-selection biases in the experiments (i.e., workers may self-select into the treatment that they
can complete tasks faster or find more enjoyable) if we adopt such a design.

To overcome the drawback of this simple design, we design our experimental treatments in a way that each treatment consists of the same number
of independent tasks {\em and} discussion tasks, such that neither treatment appears to be obviously more time-consuming or enjoyable. 
Figure~\ref{fig:exp-design} illustrates the two treatments used in our experiments.
In particular, we bundle 6 tasks in each HIT\footnote{HIT stands for Human Intelligence Task, and it refers to one unit of job on MTurk that a worker can accept to work on.}. 
When a worker accepts our HIT, she is told that there are 4 independent tasks and 2 discussion tasks in the HIT. 
There are two treatments in our experiments: in Treatment 1, workers are asked to complete 4 independent tasks followed by 2 discussion tasks, while workers in Treatment 2 first work on 2 discussion tasks and then complete 4 independent tasks.
Importantly, we do {\em not} tell workers the ordering of the 6 tasks, which helps us to minimize the self-selection biases as the two treatments look the same to workers.
We refer to the first, middle, and last two tasks in the sequence as Session 1, 2, 3 of the HIT, respectively.

Given the way we design the treatments, we can examine H1 by comparing the work quality produced in Session 1 (i.e. the first two tasks of the HIT) between the two treatments. 
Intuitively, observing higher work quality in Session 1 of Treatment 2 would imply that \peer can enhance work quality above the level of independent worker performance.
Similarly, we can test H2 by comparing the work quality in Session 2 (i.e. the middle two tasks of the HIT) between the two treatments.
H2 is supported if the work quality in Session 2 of Treatment 2 is also higher than that of Treatment 1, which would suggest that after communicating with peers, 
workers are able to produce higher quality {\em in their independent work} for the same type of tasks.
Finally, Session 3 (i.e. the last two tasks of the HIT) is used to ensure that the two treatments require similar amount of work from workers. 
\ignore{It's important to note that we have decided {\em not} to draw any causal conclusions using data collected in Session 3 in the experimental design phase, because Session 3 of the two treatments differs
on both whether workers have worked in pairs in previous tasks and whether 
 as and the data collected in these two tasks is discarded\footnote{On a side note, analyzing the data collected in the last two tasks leads to conclusions that are consistent with our findings (about H1) reported below, and including such data only strengthens our results. 
However, since we have decided not to use it in the experiment design phase,
we do not include the data in our analyses.}}

\subsection{Experimental Tasks}

We conduct our experiments on three types of tasks: image labeling, optical character recognition (OCR), and audio transcription. 
These tasks are all very common types of tasks on crowdsourcing platforms, hence experimental results on these tasks allow us to understand how \peer affects the quality of crowdwork for various kinds of typical tasks. 

\begin{figure}[t]
   \centering
   \includegraphics[width=0.8\columnwidth, keepaspectratio]{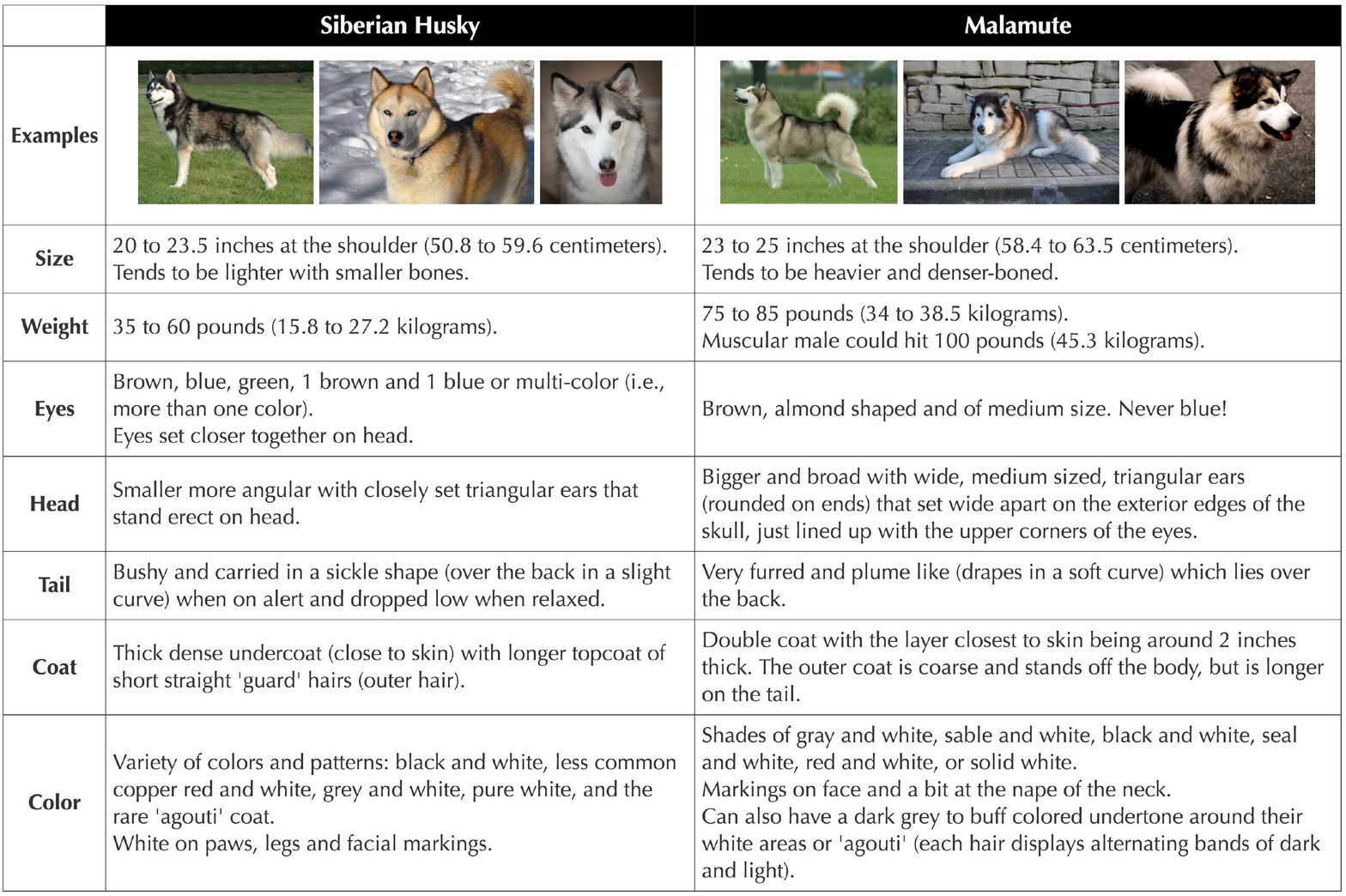}
   \vspace{-25pt}
   \caption{The instruction of the image labeling task.}
   \label{fig:dog-instruction}
   \vspace{-10pt}
\end{figure}

\begin{figure}[t]
   \centering
   \includegraphics[width=0.7\columnwidth, keepaspectratio]{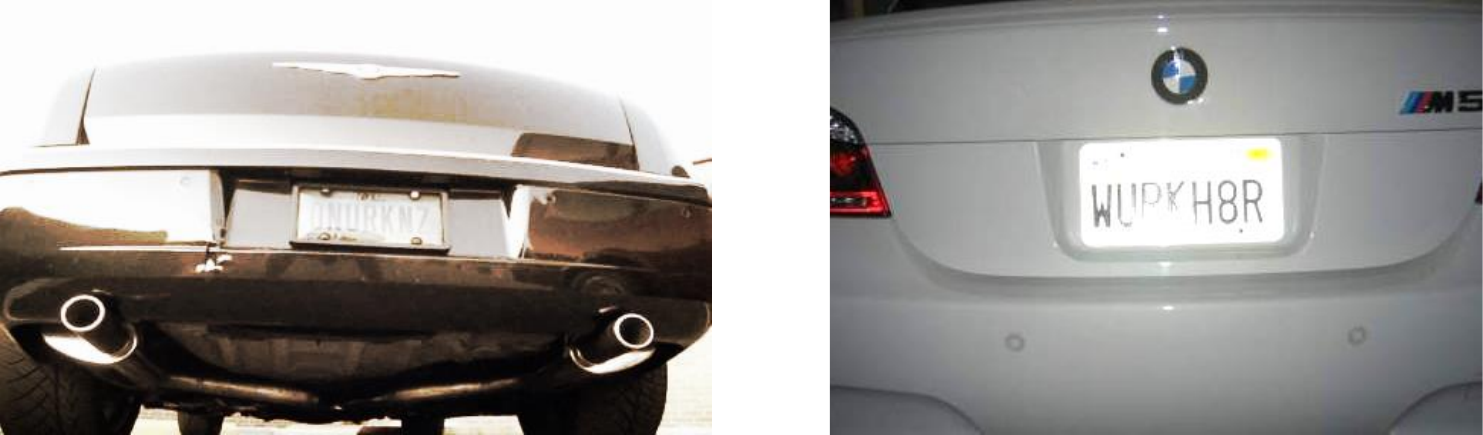}
   \vspace{-5pt}
   \caption{Examples of photos used in the OCR task. }
   \label{fig:ocr}
    \vspace{-10pt}
\end{figure}

\begin{itemize}
    \item \textbf{Image labeling.}
    In each image labeling task, we present one image to the worker and ask her to identify whether the dog in the image is a Siberian Husky or a Malamute.
    Dog images we use are collected from the Stanford Dogs dataset~\cite{KhoslaYaoJayadevaprakashFeiFei_FGVC2011}. 
    Since the task can be difficult for workers who are not familiar with dog species,
    we provide workers with a table summarizing the characteristics of each dog species, as shown in Figure~\ref{fig:dog-instruction}.
    Workers can get access to this table at anytime when working on the HIT.

    \item \textbf{Optical character recognition (OCR).}
    For the OCR task, workers are asked to transcribe vehicles' license plate numbers from photos.
    The photos are taken from the dataset provided by \citet{Shah15doubleornothing}, 
    and some examples are shown in Figure~\ref{fig:ocr}.

    \item \textbf{Audio transcription.}
    For the audio transcription task, workers are asked to transcribe an audio clip which contains approximately 5 seconds of speech. 
    The audio clips are collected from VoxForge\footnote{\url{http://www.voxforge.org}}.

\end{itemize}

Unlike in the image labeling task, we do not provide additional instructions for the OCR and audio transcription tasks. 
Indeed, for some types of crowdwork, it is difficult for requesters to provide detailed instructions. 
However, the existence of detailed task instruction may influence the effectiveness of \peer (e.g., workers in the image labeling tasks can simply discuss with their co-workers whether each distinguishing feature covered in the instruction is presented in the dog image). 
Thus, examining the effect of \peer on work quality for different types of tasks, where detailed instruction may or may not be possible,
helps us to understand whether such effect is dependent on particular elements in the design of the tasks. 
	
\subsection{Experimental Procedure}

Introducing direct communication between pairs of workers on the same tasks requires us to synchronize the work pace of pairs of workers, which is quite challenging 
as discussed in previous research on real-time crowdsourcing~\cite{Bigham10,Bernstein11}.
We address this challenge by dynamically matching workers together and sending pairs of workers to simultaneously start working on the same sequence of tasks. 

In particular, when each worker arrives at our HIT,
we first check whether there is another worker in our HIT who don't have a co-worker yet --- if yes, she will be matched to that worker and assigned to the same treatment and task sequence as that worker. 
Otherwise, the worker will be {\em randomly} assigned to one of the two treatments as well as a {\em random} sequence of tasks, and she will be asked to wait for another co-worker to join the HIT for a maximum of 3 minutes. We will prompt the worker with a beep sound if another worker indeed arrives at our HIT during this 3-minute waiting period. 
Once we successfully match a pair of workers, both of them will be automatically redirected to the first task in the HIT and they can start working on the HIT simultaneously. 
In the case where no other workers arrives at our HIT within 3 minutes, we ask the worker to decide whether she is willing to complete all tasks in the HIT on her own (and we will drop the data for the analysis but still pay her accordingly) or keep waiting for another 3 minutes and receive a compensation of 5 cents for waiting. 

For all types of tasks, we provide a base payment of 60 cents for the HIT.
In addition to the base payments, workers are provided with the opportunity to earn performance-based bonuses, that is, workers can earn a bonus of 10 cents in a task if the final answer they submit for that task is correct. 
Our experiment HITs are open to U.S. workers only, and each worker is only allowed to take one HIT for each type of tasks.

\subsection{Experimental Results}


For the image labeling, OCR, and audio transcription tasks, we obtain data from 388, 382, and 250 workers through our experiments, respectively\footnote{We have targeted to recruit around 200 workers for each treatment, leading to about 400 workers for each experiment.
However, we have encountered difficulties reaching workers of the targeted size for the audio transcription tasks, probably because workers consider the payment
to be not high enough for audio transcription tasks (we fix the payment magnitude to be the same across the three types of tasks).}. 
We then examine Hypothesis 1 and 2 separately for each type of task by analyzing experimental data collected from Session 1 and 2 in the HIT, respectively.
It is important to note that in the experimental design phase, we have decided {\em not} to include data collected from Session 3 of the HIT into our formal analyses.
This is because workers in Session 3 of the two treatments differ to each other both in terms of whether they have communicated with other workers about the work in previous tasks and 
whether they can communicate with other workers in the current tasks, making it difficult to draw any causal conclusions on the effect of \peer. However, as we will mention below, analyzing the data collected in Session 3 leads to observations that are consistent with our findings.

\subsubsection{Work Quality Metrics}

We evaluate the work quality using the notion of \emph{error}.
Specifically, in the image labeling task, since workers can only submit binary labels (i.e., Siberian Husky or Malamute), 
the error is defined as the binary classification error---if a worker provides a correct label, the error is $0$, otherwise the error is $1$.
For OCR and audio transcription tasks, since workers' answers and the ground truth answers are both strings, 
we define ``error'' as the edit distance between the worker's answer and the ground truth, divided by the number of characters in the ground truth.
Naturally, in all types of tasks, a lower rate of error implies higher work quality.

\begin{figure}[t]
   \centering
   \includegraphics[width=0.9\columnwidth, keepaspectratio]{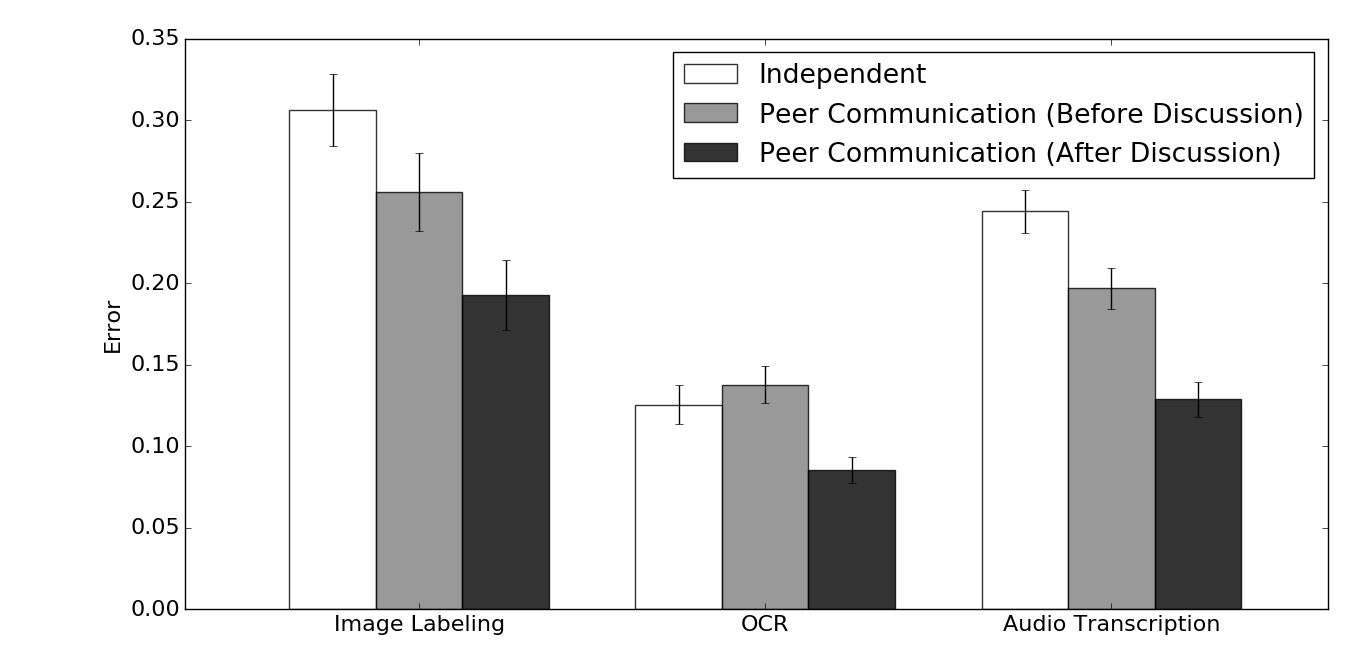}
   \vspace{-10pt}
   \caption{Examine whether workers produce higher work quality in tasks with \peer than in tasks without \peer. 
   In ``Independent'' group, we calculate workers' average error rate in Session 1 of Treatment 1 (see Figure~\ref{fig:exp-design}).
   In ``\PEER (Before Discussion)'' group, we calculate workers' average error rate in Session 1 of Treatment 2, {\em before} they communicate with co-workers about the work (i.e., for their independent answers).
   In ``\PEER (After Discussion)'' group, we calculate workers' average error rate in Session 1 of Treatment 2, {\em after} they communicate with co-workers about the work (i.e., for their final answers).
   Error bars indicate the mean $\pm$ one standard error.}
   \label{fig:h1}
   \vspace{-10pt}
\end{figure}

\subsubsection{Work Quality Improves in Tasks with \PEER}

We start with examining Hypothesis 1 by comparing work quality produced in Session 1 of the two treatments for each type of tasks.
In Figure~\ref{fig:h1}, We plot the average error rate for workers' {\em final} answers in Session 1 of Treatment 1 and 2 using white and black bars, respectively.
Visually, it is clear that for all three types of tasks, the work quality is higher after workers communicate with others about the work compared to
when workers need to complete the work on their own. We further conduct two-sample t-tests to check the statistical significance of the differences, and 
p-values for image labeling, OCR and audio transcription tasks are $2.42\times 10^{-4}$, $5.02\times 10^{-3}$, and $1.95\times 10^{-11}$ respectively, 
suggesting the improvement in work quality is statistically significant. Our experimental results thus support Hypothesis 1.

\begin{figure}[t]
   \centering
   \includegraphics[width=0.55\columnwidth, keepaspectratio]{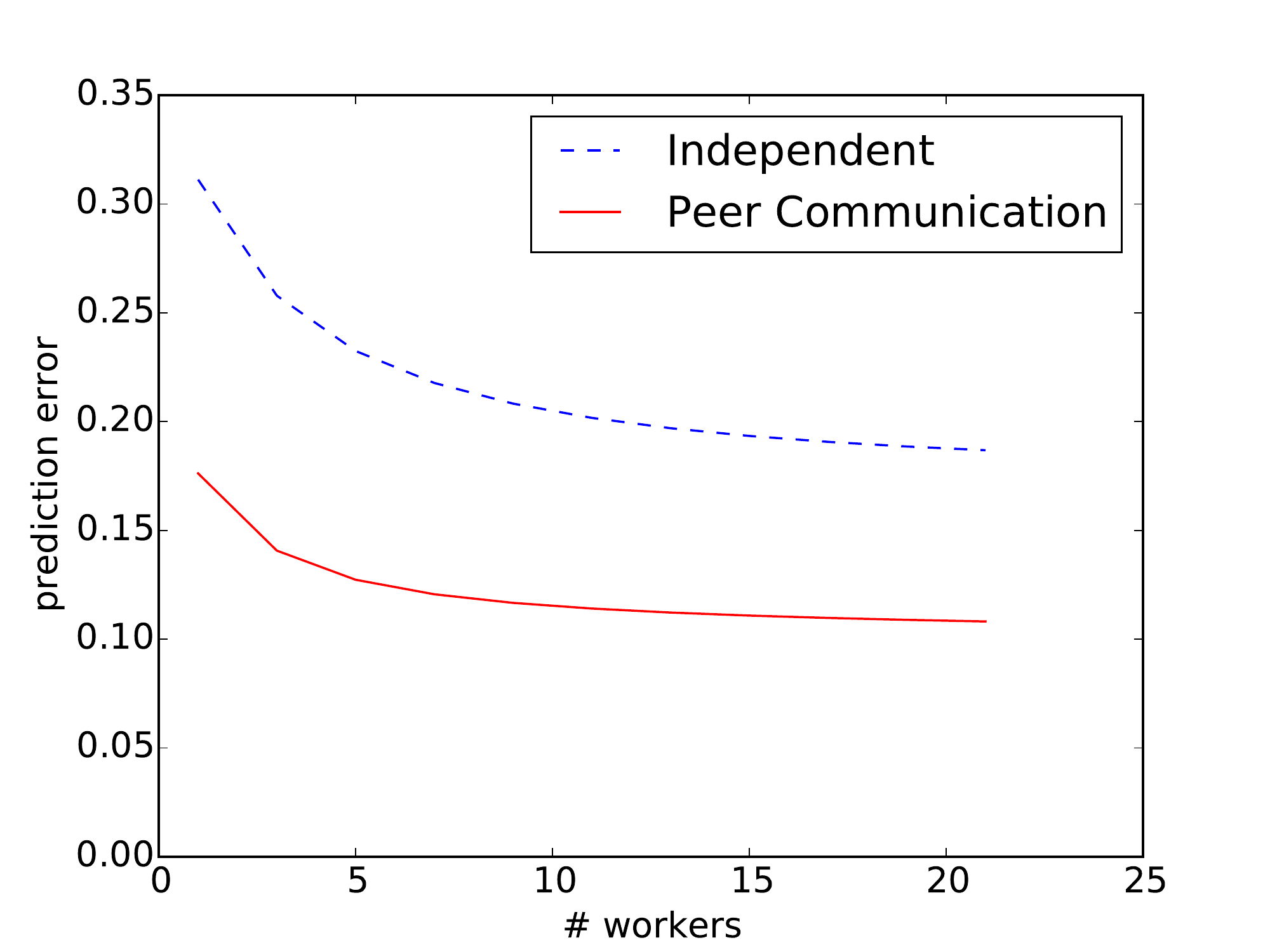}
   \vspace{-8pt}
   \caption{The aggregation error for the image labeling task after using majority voting for aggregation.}
   \vspace{-5pt}
   \label{fig:dog-aggregation}
\end{figure}

Our consistent observations on the effectiveness of \peer in enhancing the quality of crowdwork for various types of tasks indicate that enabling
direct, synchronous and free-style communications between pairs of workers who work on the same tasks might be a simple method for improving worker performance
that can be easily adapted to different contexts. 
To further highlight the advantage of \peer, we apply majority voting to aggregate the labels obtained during Session 1 of the image labeling tasks\footnote{Since there is no straight-forward way to aggregate workers' answers in the other two types of tasks, we only perform the aggregation for image labeling tasks.},
and the results are presented in Figure~\ref{fig:dog-aggregation}. 
The X-axis represents the number of workers from whom we elicit labels for each image, and the Y-axis represents the prediction error (averaged across all images) of the aggregate label decided by the majority voting rule.
As we can see in the figure, the aggregation error using labels obtained from tasks with \peer greatly outperforms the aggregation error using labels from independent work.
Moreover, in independent tasks, a majority of workers provide incorrect labels for approximately 20\% of the images (therefore, the prediction error converges to near 20\%) while in tasks with \peer, this aggregated error reduces to only around 10\%. These results reaffirm the superior quality of data collected through tasks with \peer.

A natural question one may ask then is {\em why} work quality improves in tasks with \peer. 
One possible contributing factor is the {\em social pressure}, that is, workers may put more effort and thus produce higher work quality in the tasks, 
simply because they know that they are working with a co-worker on the same task and are going to discuss with the co-worker about the task. 
Another possibility is that {\em constructive conversations} between workers enable effective knowledge sharing and lead to the improvement in work quality. 
To get a better understanding on the role of these two factors on influencing work quality in tasks with \peer, we conduct a few additional analyses.

\paragraph{The impacts of social pressure.}
First, we look into whether workers behave differently when they are working on their {\em independent} answers in tasks with \peer and when they are working on tasks without \peer. 
Intuitively, if workers are affected by social pressure in tasks with \peer, they may spend more time on the tasks and possibly produce work of higher quality even
at the stage when they are asked to work on the tasks on their own before communicating with their co-workers. 
Table~\ref{tab:time:before} summarizes the amount of time workers spend on tasks in Session 1 of Treatment 1, and
on Session 1 of Treatment 2 when they work on their independent answers. We find that, overall, knowing the existence of a co-worker who works on the same task makes workers
spend more time on the task on their own, though the differences are not always significant. 
In addition, we plot the average error rate for workers' independent answers in Session 1 of Treatment 2 as gray bars in Figure~\ref{fig:h1}. 
Comparing the white and gray bars in Figure~\ref{fig:h1}, we find that workers only improve the quality of their independent answers significantly in the audio transcription tasks when they know the existence of a co-worker ($p=0.010$).
Together, these results imply that workers in tasks with \peer might be affected by the social pressure to some degree, but social pressure is likely not
the major cause of the work quality improvement in tasks with \peer.


\begin{table}[t]
\centering
\begin{tabular}{c|ccc}
\toprule
\textbf{Task Type} & \textbf{Treatment 1} & \textbf{Treatment 2 (before discussion)} & \textbf{p-values}\\
\midrule
Image labeling & 16.34 (0.78) & 21.43 (0.97) & 4.165$\times10^{-5}$\\
OCR & 23.63 (1.00) & 26.13 (1.13) & 0.099\\
Audio transcription & 55.91 (3.38) & 59.90 (2.62) & 0.353\\
\bottomrule
\end{tabular}
\vspace{5pt}
\caption{Comparison of the average amount of time (in seconds) workers spend on independently working on a task in Session 1 of Treatment 1 and 2; mean values and standard errors (in parentheses) are reported. Two-sample t-tests are used to examine whether the differences are statistically significant, and p-values are reported in the last column.}
~\label{tab:time:before}
\vspace{-25pt}
\end{table}

\ignore{To examine whether social pressure is a factor to improve the work quality,
we compare the following: (1) workers' average error on the first two tasks in treatment 1 and (2) workers' average error on the initial answers (before discussion) of the first two tasks in treatment 2. 
With this comparison, we can remove the effect of the discussion and see whether just telling workers that they are going to discuss their answers with others would motivate them to work harder.
The results are shown in Figure~\ref{fig:before-discussion}.
For the audio transcription task, it seems having social pressure encourages workers to generate better quality of work (p-value is $0.01$ for two-sample t-test). However, we do not observe statistically significant differences in the other two tasks. Overall, the results on whether social pressure encourages higher-quality work seems to be task-dependent.}

\ignore{
\begin{figure}[ht]
   \centering
   \includegraphics[width=0.9\columnwidth, keepaspectratio]{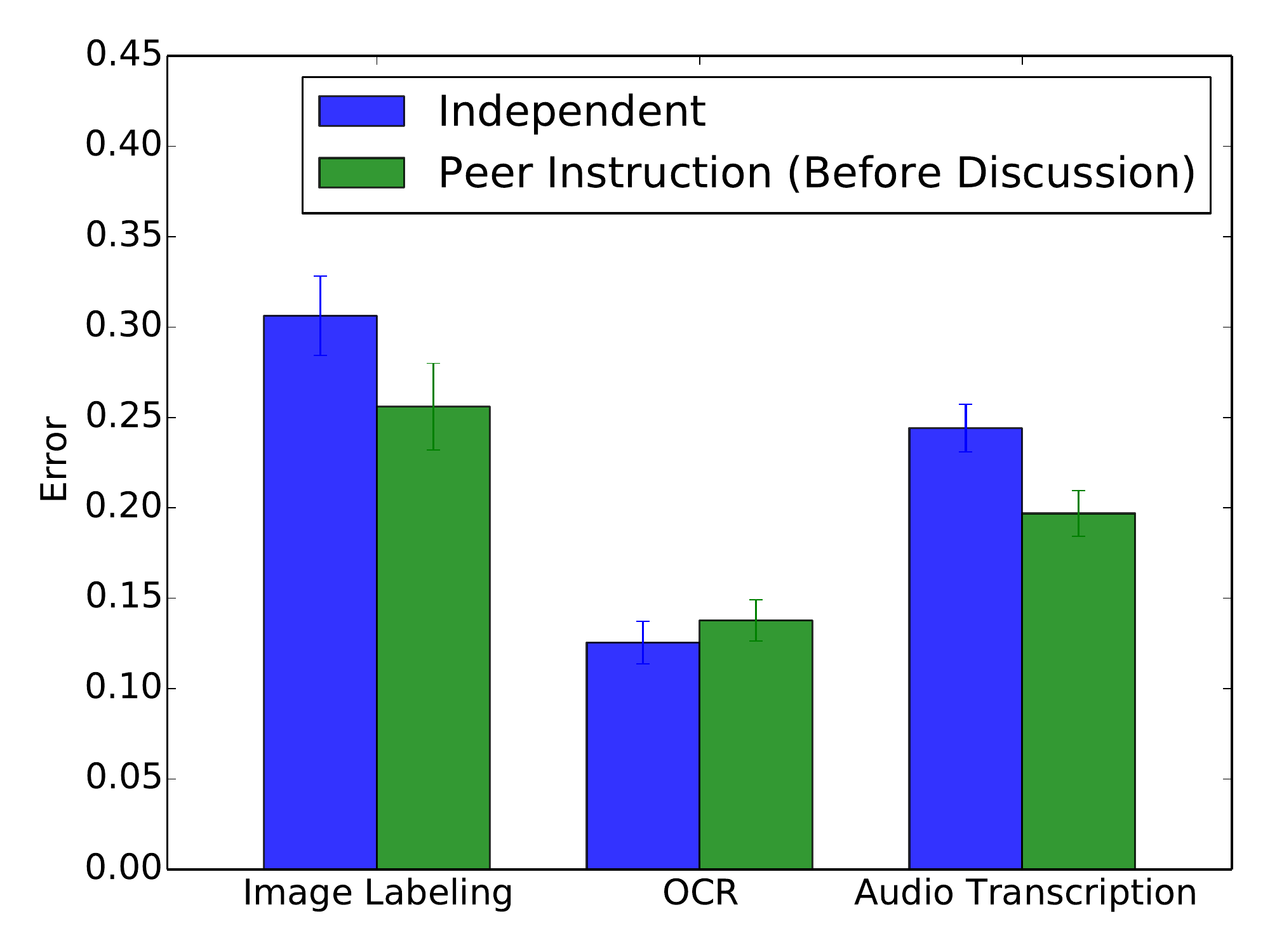}
   \caption{Examine whether social pressure improves workers' performance.
   In ``Independent'' group, we calculate workers' average error in task 1 and 2 of treatment 1 (see Figure~\ref{fig:exp-design}).
   In ``Peer Instruction (Before Discussion)'' group, we calculate workers' average error on their initial answers (before discussion) in task 1 and 2 of treatment 2.}
   \label{fig:before-discussion}
\end{figure}
}

\begin{figure}[t]
   \centering
   \includegraphics[width=\columnwidth, keepaspectratio]{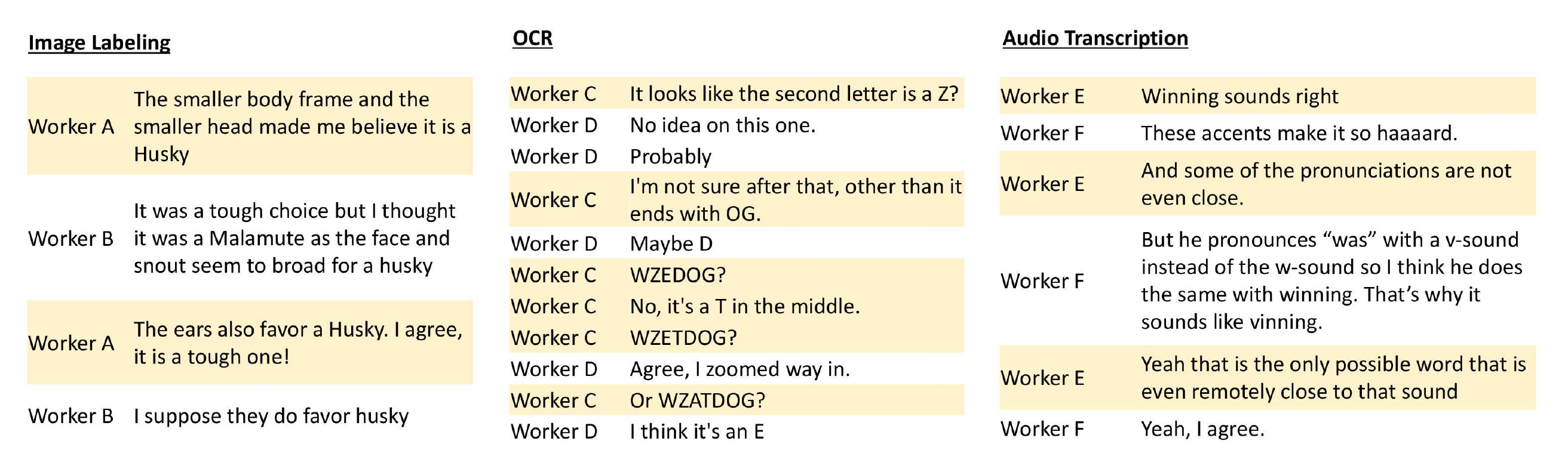}
   \vspace{-25pt}
   \caption{Examples of chat logs.
   }
   \vspace{-15pt}
   \label{fig:chat-log}
\end{figure}

\paragraph{The impacts of constructive conversations.}
Next, we examine whether the conversations between co-workers, by itself, help workers in tasks with \peer to improve their work quality. 
We thus compare the quality of workers' independent answers before discussion (gray bars in Figure~\ref{fig:h1}) and their final answers after discussion 
(black bars in Figure~\ref{fig:h1}) in Session 1 of Treatment 2. 
We find that workers in tasks with \peer submit final answers of higher quality after discussion than their independent answers before discussion.
We further conduct paired t-tests on worker's error rate before and after discussion for tasks in Session 1 of Treatment 2, and test results show that the difference is statistically significant for all three types of tasks ($p=5.09\times 10^{-4}, 1.51\times 10^{-9}$ and $3.62\times 10^{-8}$ for image labeling, OCR, and audio transcription tasks, respectively). 
In fact, we can also reach the same conclusion if we conduct a similar analysis for the work quality produced before and after discussions in Session 3 (i.e., the last two tasks) of Treatment 1. That is to say, the communication between co-workers about the same piece of work consistently leads to a significant improvement in work quality. 

To gain some insights on what workers have communicated with each other during the discussion, we show a few representative examples of chat logs in Figure~\ref{fig:chat-log}.
We find that workers are mostly engaged in constructive conversations in the discussions. In particular, workers not only try to explain to each other the reasons why they come up with their independent answers and deliberate on whose answer is more convincing (as shown in the example for the image labeling tasks), but they also try to jointly work on the tasks together (as shown in the example for the OCR tasks). 
Throughout the conversations, workers communicate on their confidence about their answer (e.g., ``I'm not sure after that...'') as well as their strategies for solving the tasks (e.g., ``he pronounces `was' with a v-sound instead of the w-sound''). 
Note that much of the discussions as shown in Figure~\ref{fig:chat-log} can hardly be possible without allowing workers to directly interact and exchange information with each other in real time, which implies the necessity of direct, synchronous, free-style interactions between workers in crowdwork.  

To briefly summarize, we have consistently found that enabling \peer among pairs of workers can enhance the work quality for various types of tasks above the level of independent worker performance, which can be partly attributed to the social pressure brought up by the \peer process, but is mostly due to the constructive conversations between workers about the work. 
These results indicate that introducing \peer in crowdwork can be a simple, generalizable approach to enhance work quality.

\subsubsection{Effects of \PEER on Work Quality in Future Tasks}
\label{exp1:H2}

We now move on to examine our Hypothesis 2: compared to workers who have never been involved in \peer, do workers who have participated in tasks with \peer continue to produce work of higher quality in future tasks of the same type, even if they need to complete those tasks on their own? 
In other words, is there any ``spillover effect'' of \peer on the quality of crowdwork, such that \peer can be used as a ``training'' method to enhance workers' independent work quality in the future?

\begin{figure}[t]
   \centering
   \includegraphics[width=0.52\columnwidth, keepaspectratio]{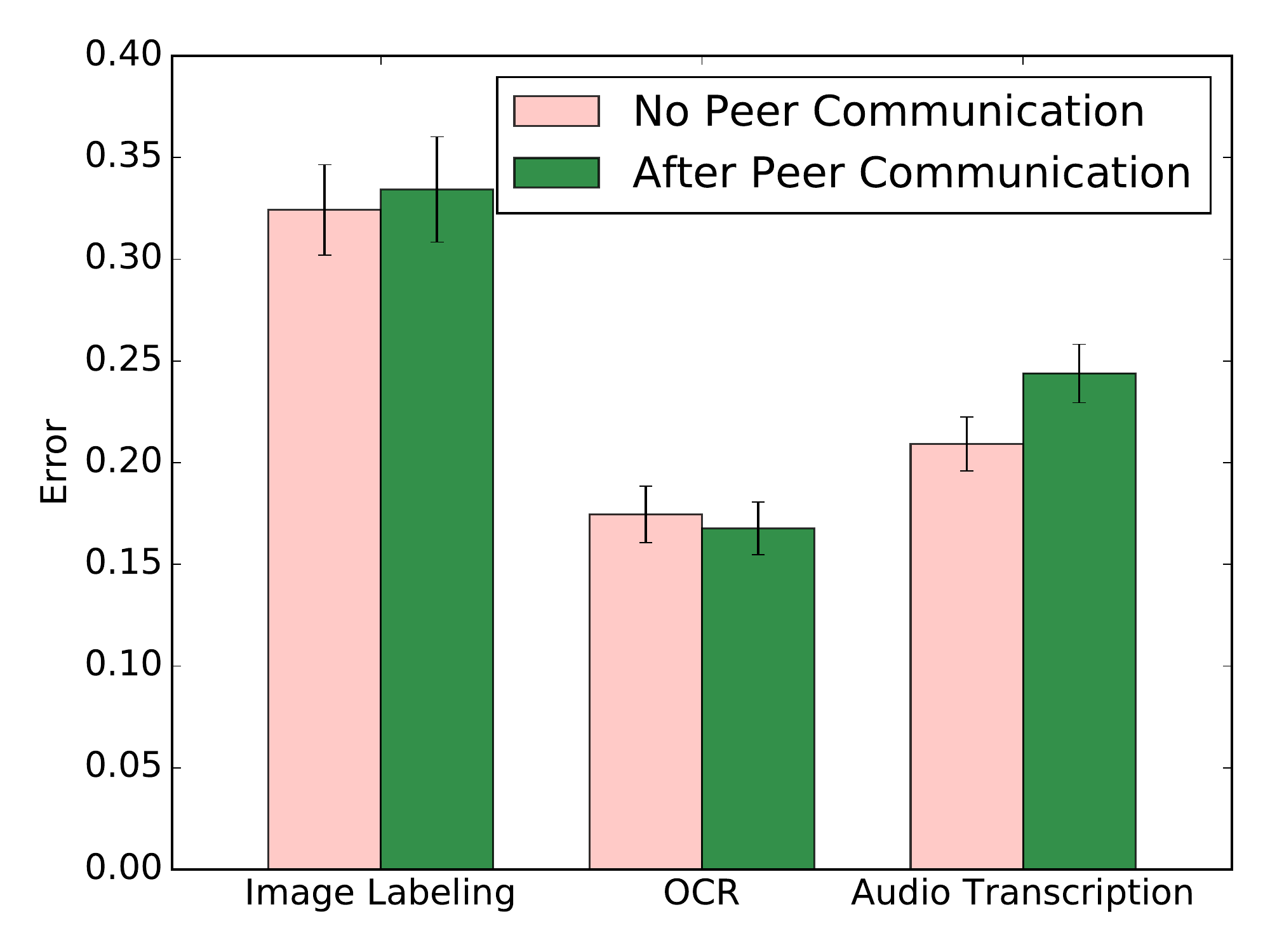}
   \vspace{-10pt}
   \caption{Examine whether the work quality in future tasks of the same type increases after workers participating in tasks with \peer. 
   In ``No Peer Communication'' group, we calculate workers' average error rate in Session 2 of Treatment 1.
   In ``After Peer Communication'' group, we calculate workers' average error rate in Session 2 of Treatment 2.
   Error bars indicate the mean $\pm$ one standard error.}
   \label{fig:h2}
   \vspace{-15pt}
\end{figure}

To answer this question, we compare the work quality produced in Session 2 (i.e., the middle two independent tasks) 
of the two treatments for all three types of tasks, and results are shown in Figure~\ref{fig:h2}. 
As we can see in the figure, 
there are no significant differences in work quality between treatments, indicating that after participating in tasks with \peer, workers are {\em not} able to maintain a higher level of quality when they complete tasks of the same type on their own in the future. 
Therefore, our observations in Session 2 of the two treatments do not support Hypothesis 2. To fully understand whether and when Hypothesis 2
can be supported, we continue to conduct a set of follow-up experiments, which we will describe in detail in the next section.  

\ignore{
The results are not particularly surprising. For most of the tasks in crowdsourcing markets, there are not a lot of new concepts to learn. 
While peer instruction can improve the quality for each individual task, workers might not be able to learn new concepts during discussion phases that are applicable to other similar tasks.
Moreover, even if workers can learn useful new concepts during discussion, we might need to give workers more time for them to grasp the concept and apply them in tasks. 
Identifying tasks that can benefit from ``teaching'' workers to better solve tasks and examining whether peer instruction can help for those tasks would be an interesting future direction.}

\label{sec3}

\section{Experiment 2: When Does Peer Communication Affect Quality of Independent Work in Future Tasks?} 
The results of our previous experiment do not support Hypothesis 2, i.e., after participating in tasks with \peer, workers do not produce work of higher quality in tasks of the same type when working independently. 
This is in contrast with the empirical findings of peer instruction in educational setting, despite that the procedure of \peer is adapted from peer instruction.
We conjecture that two factors may have contributed to this observed difference. 


First, for peer instruction, concepts covered in the post-tests 
(e.g., when students answer the test questions on their own after the instruction ends) 
are often the {\em same} as concepts discussed during the peer instruction process in class.
Therefore, knowledge learned from the peer instruction process can be directly transferred to post-tests. 
This is not necessarily true for peer communication in crowdwork---for example, when workers are asked to complete a sequence of tasks to identify Siberian Husky and Malamute, 
it is possible that the distinguishing feature for the dog in one task is its eyes while the distinguishing feature for the dog in another task is its size,
making the knowledge that workers possibly have learned in tasks with \peer not always useful on future tasks that are somewhat unrelated.

In addition, as we have discussed in Section~\ref{sec:pi}, compared to the standard peer instruction procedure, 
we remove the last step (see Figure~\ref{fig:PI-workflow}) where the requester provides expert feedback to workers after reviewing workers' final answers in the \peer process\footnote{This step would be equivalent to instructor reviewing students' final responses and providing more explanation as needed in peer instruction.}  
due to the low availability of expert feedback. 
It is thus possible that worker's quality improvement in future independent work 
can only be obtained when additional expert feedback is provided after \peer.


Therefore, in this section, we conduct an additional set of experiments to examine whether these two factors have impacts on the effectiveness of \peer as a tool for training workers, and thus
seek for a better understanding on whether and when \peer can affect workers' independent performance in the future.

\subsection{Experimental Tasks} 

\begin{figure}[t]
   \centering
   \begin{subfigure}[b]{0.9\textwidth}
   \centering
   \includegraphics[width=0.9\columnwidth, keepaspectratio]{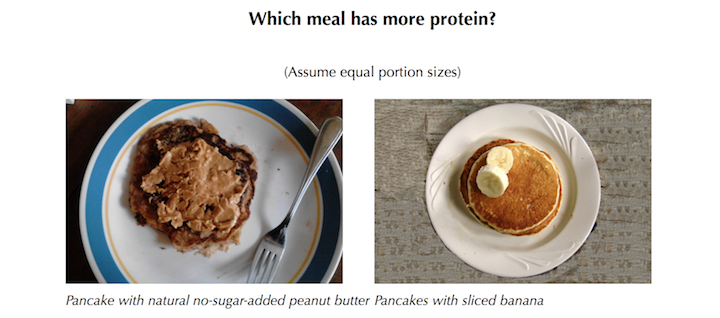}
   \caption{Example of a nutrition analysis task}~\label{fig:nutrition:task}
   \end{subfigure}
   \begin{subfigure}[b]{1\textwidth}
   \centering
   \includegraphics[width=\columnwidth, keepaspectratio]{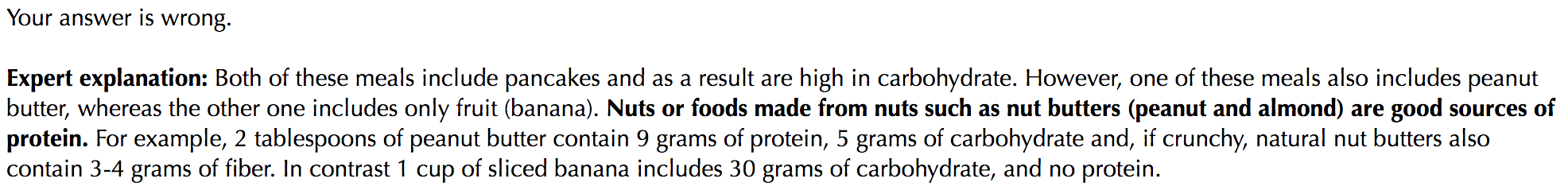}
   \caption{Expert feedback for the above nutrition analysis task}~\label{fig:nutrition:explanation}
   \end{subfigure}
   \vspace{-20pt}
   \caption{Example of a nutrition analysis task and the expert explanation associated with it.}\label{fig:nutrition}
   \vspace{-10pt}
\end{figure}

In this study, we use nutrition analysis tasks, provided in the work by ~\citet{burgermaster2017role}, in the experiments.
In each nutrition analysis task, we present a pair of photographs of mixed-ingredient meals to workers. 
Workers are asked to identify which meal in the pair contains a higher amount of a specific macronutrient (i.e., carbohydrate, fat, protein, etc.). 
To help workers figure out the main ingredients of the meals in each photograph, 
we also attach a textual description with each photograph. 
Figure~\ref{fig:nutrition:task} shows an example of a nutrition analysis task.

We choose to use the nutrition analysis tasks for two reasons. 
First, each nutrition analysis task is associated with a ``topic,'' which is the key concept underlying the task. 
For example, the topic for the task shown in Figure~\ref{fig:nutrition:task} is that nuts (contained in peanut butter) are important sources of proteins. 
Knowing the topic of each task, we can then place tasks of the same topic subsequently and examine whether, after participating in tasks with \peer, workers improve their independent work quality on {\em related} tasks (i.e., tasks that share the same underlying concept). 
Second, we have access to expert explanations for each nutrition analysis task (see Figure~\ref{fig:nutrition:explanation} for an example), 
which allows us to test whether \peer has to be combined with expert feedback to influence worker's independent performance in future tasks.

We would like to note that the underlying concepts and expert feedback are often hard to obtain in crowdsourcing tasks,
since requesters do not know the ground truth. 
The purpose of this follow-up study is to provide us better insights on under what conditions can \peer be an effective tool for training workers.

\subsection{Experimental Design}

We explore whether \peer can be used to train workers 
when combined with the two factors we discuss above: whether tasks are conceptually similar
(i.e., whether future tasks are \emph{related} to tasks where \peer is enabled), 
and whether \emph{expert feedback} is provided at the end of \peer.

In particular, we aim to answer whether \peer is effective in training workers when 
(a) tasks are conceptually similar but no expert feedback is given at the end of \peer,
(b) tasks are not conceptually similar but expert feedback is given at the end of \peer, or 
(c) tasks are conceptually similar and expert feedback is given at the end of \peer.
For both the second and the third question, if the answer is positive, a natural question then is whether the improvement on independent work quality in future tasks is attributed to the expert feedback or the \peer procedure.

Corresponding to these three questions, we design three sets of experiments. 
All the experiments share the same structure as the experiments we have designed in the previous section.
That is, we include two treatments in each experiment, where Treatment 1 contains 4 independent nutrition analysis tasks followed by 2 discussion tasks and Treatment 2 contains 2 discussion tasks followed by 4 independent tasks. 
We highlight the differences in the design of these three experiments in the following.

\subsubsection{Experiment 2a}
Different from that in experiments of Section 3, in this experiment, tasks within a HIT are {\em not} randomly selected. 
Instead, for both treatments in this experiment, tasks in Session 1 are randomly picked, 
while tasks in Session 2 are selected from the ones that share the {\em same} topics as tasks in Session 1. 
This experiment is designed to understand whether \peer can lead to better independent work quality in {\em related} tasks in the future. 
Naturally, if workers are able to achieve better performance in Session 2 of Treatment 2 compared to that in Session 2 of Treatment 1, 
we may conclude that workers can improve their independent work quality after participating in tasks with \peer, 
but only for those related tasks that share similar concepts as the tasks that they have discussed with other workers.      

\subsubsection{Experiment 2b}
The main difference between this experiment and experiments of Section 3 is the presence of expert feedback. 
Specifically, for all discussion tasks in this experiment, after workers submit their final answers, 
we will display extra information to workers which includes a feedback on whether the worker's final answer is correct, 
and an expert explanation on why the worker's answer is correct or wrong (see Figure~\ref{fig:nutrition:explanation} for an example). 
Workers are asked to spend at least 15 seconds reading this information before they can proceed to the next page in the HIT. 
Note that in this experiment, the tasks included in a HIT are {\em randomly} selected. Comparing worker's performance in Session 2 of the two treatments in this experiment, 
thus, inform us on whether the addition of expert feedback at the end of the \peer procedure will lead to improvement on independent work quality in future tasks, 
which may or may not be related to the tasks for which \peer is enabled.

\subsubsection{Experiment 2c}
Our final experiment is the same as Experiment 2b, except for one small difference---tasks included in Session 2 have the same topics as tasks in Session 1. 
This experiment then allows us to understand whether workers' independent work quality improves after they participate in tasks with \peer, 
when expert feedback is combined with peer communication {\em and} future tasks are related to the tasks with \peer. 

Our experiments are open to U.S. workers only, and each worker is allowed to participate in only one experiment.

\subsection{Experimental Results}

\begin{figure}[t]
   \centering
   \begin{subfigure}[b]{0.32\textwidth}
   \centering
   \includegraphics[width=\columnwidth, keepaspectratio]{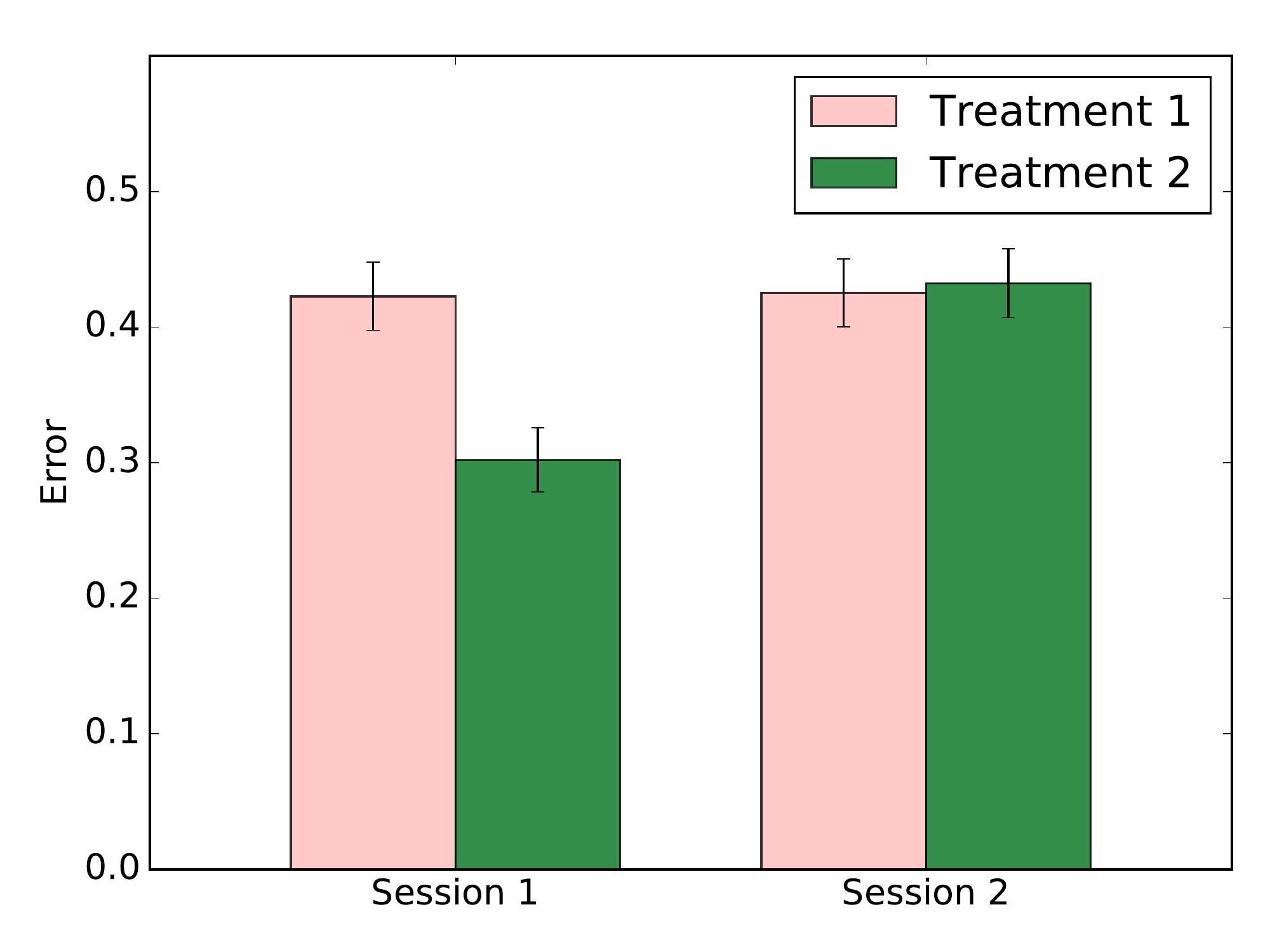}
   \caption{Experiment 2a results}~\label{fig:nutrition:2a}
   \end{subfigure}
   \begin{subfigure}[b]{0.32\textwidth}
   \centering
   \includegraphics[width=\columnwidth, keepaspectratio]{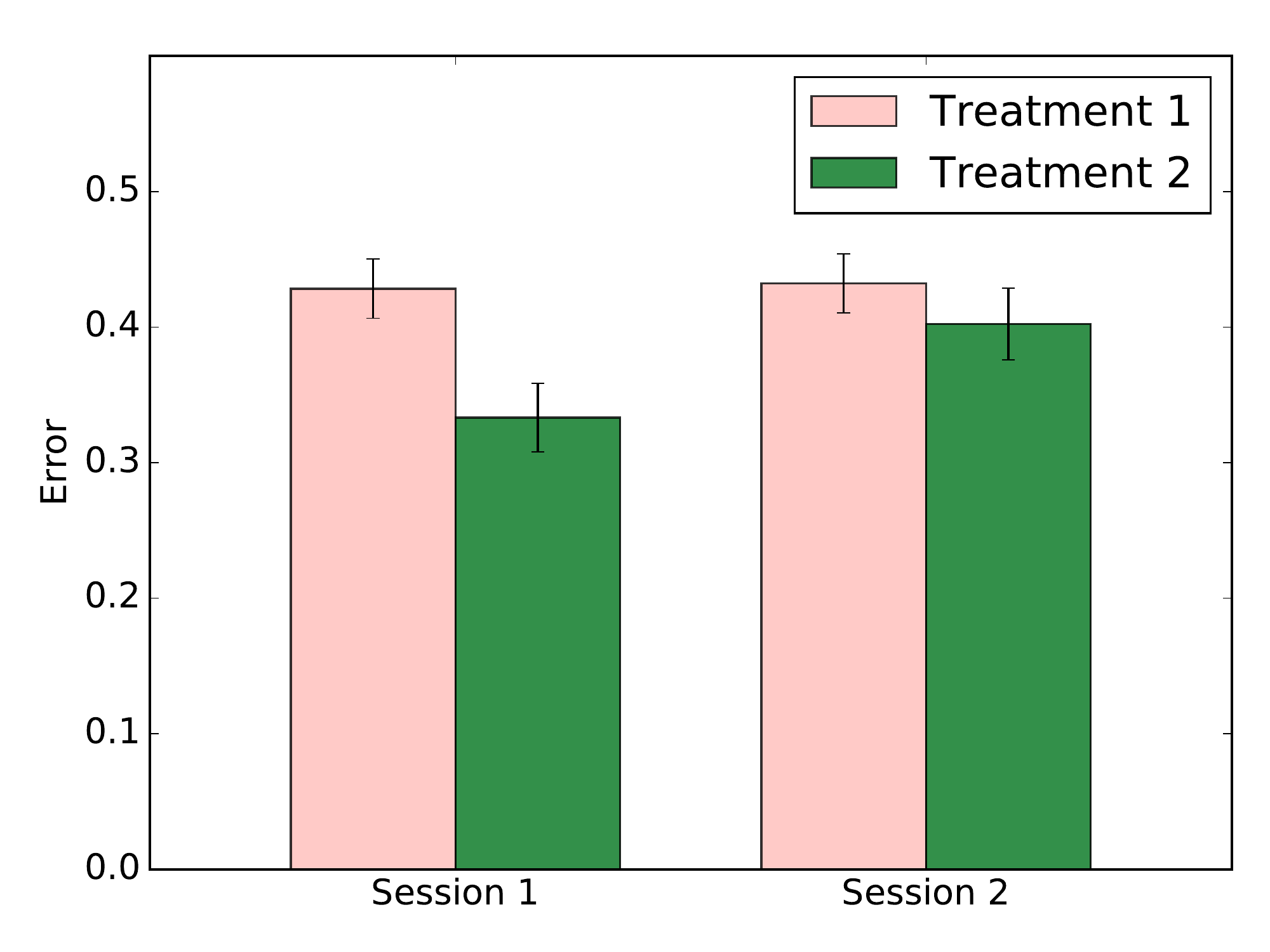}
   \caption{Experiment 2b results}~\label{fig:nutrition:2b}
   \end{subfigure}
   \begin{subfigure}[b]{0.32\textwidth}
   \centering
   \includegraphics[width=\columnwidth, keepaspectratio]{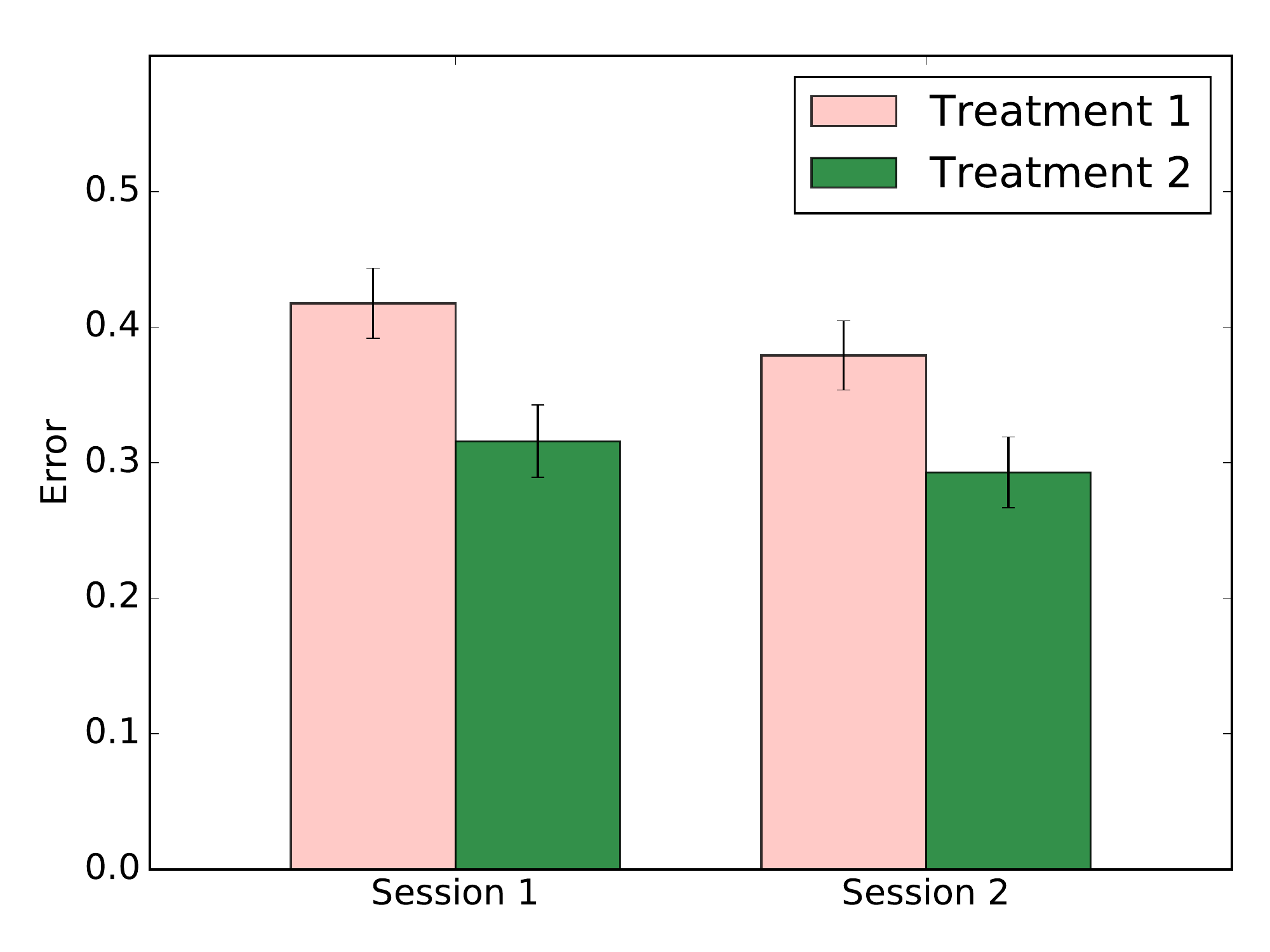}
   \caption{Experiment 2c results}~\label{fig:nutrition:2c}
   \end{subfigure}
   \vspace{-15pt}
   \caption{Comparison of workers' average error rates in the three sets of experiments in which we examine whether workers improve the quality of independent work after they
  participate in tasks with \peer. }\label{fig:exp2}
   \vspace{-10pt}
\end{figure}

In total, 386, 432 and 334 workers have participated in Experiments 2a, 2b and 2c, respectively. 
Figure~\ref{fig:exp2} shows the results on the comparison of work quality in the two treatments for all three experiments. 

First, we notice that in all three experiments, there are significant differences in the work quality between the two treatments for tasks in Session 1, which reaffirms
our findings that workers significantly improve their performance in tasks with \peer compared to when they work on the tasks by themselves (p-values for two-sample t-tests on workers' error rates in Session 1 are $4.779\times10^{-4}$, 0.005, and 0.007 for Experiments 2a, 2b, 2c, respectively).  

Furthermore, for tasks in Session 2, we find that workers don't exhibit much difference in their work quality for tasks in Session 2 between the two treatments in Experiment 2a or 2b (p-values for two-sample t-tests on workers' error rates in Session 2 are 0.844 and 0.384 for Experiment 2a and 2b, respectively), but there is a significant difference for the work quality in Session 2 between the two treatments in Experiment 2c ($p=0.019$). Together with our findings in Section~\ref{exp1:H2}, these results imply that simply enabling direct communication between pairs of workers who work on the same microtasks does {\em not} help worker to improve their independent work quality in future tasks, in regardless of whether those future tasks share related concepts to the tasks that they have discussed with co-workers. In addition, simply providing expert feedback after the \peer procedure can {\em not} enhance worker's future independent performance on some randomly selected tasks of the same type, either. Nevertheless, it seems that \peer, when combined with expert feedback, can lead to improved independent work quality on future tasks that are conceptually related to the tasks where \peer takes place.

One may wonder why \peer, by itself, can hardly influence workers' independent work quality in future related tasks (i.e., results of Experiment 2a), but can influence workers' independent work quality when it is combined with expert feedback (i.e., results of Experiment 2c). We provide two explanations to this. First, by looking into the chat logs, we find that while workers often discover the underlying concept for a task through the \peer procedure, their understandings on that concept are often {\em context-specific} and are not always generalizable to a different context. For example, given two tasks that share the same topic of ``nuts are important sources of protein,'' a pair of workers might have successfully concluded in one task through discussion that peanut butter in one meal has more protein than banana in the other meal. However, when they are asked to complete the related task on their own, they are facing a choice between a meal with peanut butter and a meal with cream cheese, for which their knowledge about peanut butter that they have learned previously are not entirely applicable. In other words, worker may lack the ability to generalize the transferrable knowledge from concrete tasks through a short period of \peer (e.g., within 2 microtasks).

Perhaps more importantly, we argue that the improvement in worker's independent work quality after participating in tasks with \peer and expert feedback is largely due to the existence of expert feedback, rather than the \peer procedure. 
To show this, we conduct a follow-up experiment with the same two-treatment design as Experiment 2c, 
except that we provide expert feedback on the first two independent tasks of Treatment 1. 
In this way, comparing the work quality produced in Session 2 of the two treatments, 
we can understand whether the \peer procedure provides any additional boost to worker's independent performance beyond the improvement brought up by expert feedback. 
Our experimental results on 494 workers give a negative answer---on average, workers' error rates in Session 2 of Treatment 1 and 2 are 22.7\% and 27.3\%, respectively, and the difference is not statistically significant ($p=0.093$). 

Overall, our examinations on the effects of \peer on workers' independent work quality in future tasks suggest a limited impact. In other words, \peer may not be a very effective approach to ``train'' workers towards a higher level of independent performance, at least when workers are working on microtasks for a relatively short period of time.

\section{Discussion}
wIn this paper, we have studied the effects of direct interactions between workers in crowdwork, and in particular,
we have explored whether introducing \peer in tasks can enhance work quality in those tasks as well as improving workers'
independent work performance in future tasks of the same type. Our results indicate a robust improvement in work quality
when pairs of workers can directly communicate with each other, and such improvement is consistently observed
across different types of tasks. On the other hand, we also find that allowing workers to communicate with each other in some tasks may have limited impacts on improving
workers' independent work performance in tasks of the same type in the future.

\subsection{Design Implications}
Our consistent observations on the improvement of work quality in tasks with \peer indicate an alternative way of organizing
microtasks in crowdwork: instead of having workers solving microtasks independently, practitioners may consider the possibility
of systematically organizing crowd workers to work in pairs and enabling direct, synchronous and free-style interactions between pairs of workers
to enhance the quality of crowdwork. In some sense, our results suggest the promise and potential benefits of ``working in pairs'' as a new baseline approach
to organize crowdwork. On the other hand, introducing \peer in crowdwork also creates the complexity for requesters to synchronize the work pace
of different workers. Thus, practitioners may need to carefully deliberate on the trade-off between quality improvement brought up by \peer and extra costs of synchronizing
before they implement one specific way to organize their crowdwork.  

It is worthwhile to mention that while our experimental results show the advantage of introducing \peer in crowdwork for many different types of tasks, we can not rule out the
possibility that for some specific type of tasks, \peer may not be helpful or even be harmful. Previous studies have reported phenomenon like \emph{groupthink}~\cite{janis1982groupthink} where communication
may actually hurt the individual performance. Therefore, more experimental research is needed to thoroughly understand the relationship between the property of tasks and whether
\peer would be helpful for those tasks.

\subsection{Limitations}
While our results are overall robust and consistent,
our specific experimental design and choice of tasks imply a few limitations.

First, our experiments only span for a short period of time (i.e., six microtasks), 
and workers can only communicate with each other in two microtasks. 
This short period of interactions could be a bottleneck for workers to really \emph{learn} the underlying concepts or knowledge that is the key for workers to improve their independent
performance. Indeed, in the educational settings, students are often involved in a course that spans for hours or even months, so their improved learning performance in courses with peer instruction
could be attributed to repeated exposure to the peer instruction process.
In this sense, our observation that \peer is not an effect tool for training workers could be simply due to this short interactions.
Thus, it is an interesting future direction to explore the long term impacts of \peer for crowdwork.

Moreover, our current experiments focus exclusively on microtasks, thus it is unclear
whether our results observed in this study can be generalized to more complex tasks.
In particular, many previous work have shown that implicit worker interactions in the form of workers receiving feedback from other workers or reviewing
other workers' output can be an effective method for training workers towards better independent performance.
We conjecture that our conclusion on \peer being not a very effective training method is somewhat limited by the nature of microtasks,
and examining the effectiveness of \peer for more complex tasks is a direction that worths further study.

\subsection{Future Work}
In addition to many interesting future directions we have discussed above, there are a few more topics that we are particularly interested in exploring in the future.

First, our current study focuses on studying \peer between {\em pairs} of workers in crowdwork. Can we generalize interactions involving more than two workers? 
How should we deal with additional complexities, such as social loafing~\cite{Latan79}, when there are more than two workers involved in the communication?
It would be interesting to explore the roles of communications in multi-worker collaborations for crowdwork.

Second, in this study, we focused on implementing the component of worker interactions.
However, in peer instruction in education, instructor intervention has a big impact on student learning.
It is natural to ask, can we further improve the quality of crowdwork through requester intervention?
For example, if most workers already agree on an answer, there is a good chance the answer is correct, 
and therefore the requester can intervene and skip the discussion phase to improve efficiency. 
In general, can the requester further improve the quality of work by dynamically taking interventions in \peer process, 
e.g., by deciding whether a discussion is needed or even modify the pairing of workers based on the previous discussion?

\ignore{Moreover, our current discussion is limited to one-to-one interactions between pairs of workers.
Can we generalize interactions involving more than two workers? 
How should we deal with additional complexities, such as social loafing~\cite{Latan79} when there are more than two workers involved in the communication?
It would be interested to explore the roles of communications in multi-worker collaborations for crowdwork.}


\section{Conclusion}
In this paper, we have explored how introducing \peer---direct, synchronous, free-style interactions between workers---affects crowdwork.
In particular, we adopt the workflow of peer instruction in educational settings and examine the effects of one-to-one interactions between pairs of workers working on the same microtasks. 
Experiments on Amazon Mechanical Turk demonstrate that adopting \peer significantly increases the quality of crowdwork over the level of independent worker performance,
and such performance improvement is robust across different types of tasks. 
On the other hand, we find that participating in tasks with \peer only leads to improvement in workers' independent tasks in future tasks of the same type, if
expert feedback is provided at the end of the \peer procedure and future tasks are conceptually related to the tasks where \peer take places.
However, the improvement is likely caused by the expert feedback rather than by \peer. 
Overall, these results suggest that \peer, by itself, may not be an effective method to train workers towards better performance, at least for typical microtasks on crowdsourcing platforms. 
 

\section*{Acknowledgments}
We thank all the crowd workers who participated in the experiments to make this work possible.

{
\bibliographystyle{plainnat} 
\bibliography{my,library,qcp}
}

\end{document}